\documentclass{ctr_summer}

\usepackage{ctrfont}
\usepackage{natbib}
\usepackage{comment}
\usepackage{longtable,tabularx}
\usepackage{threeparttable}
\usepackage{setspace}
\usepackage{booktabs}
\usepackage{mathtools}
\usepackage{bm}
\usepackage{graphicx}
\DeclareGraphicsExtensions{.jpg,.jpeg,.pdf,.png,.mps,.eps,.ps}

\usepackage{psfrag}
\usepackage{epsfig}
\usepackage{amsmath,rotating}
\usepackage{dashrule}
\usepackage{undertilde}

\usepackage{url}

\usepackage{xcolor}
\usepackage{tikz}

\definecolor{magenta2}{RGB}{255,0,255}
\tikzstyle{longdashed}=                  [dash pattern=on 6pt off 2pt]
\tikzstyle{dashdotdot}=              [dash pattern=on 4pt off 2pt on \the\pgflinewidth off 1pt on \the\pgflinewidth off 2pt]
\DeclareRobustCommand{\redsolid}{\raisebox{2pt}{\tikz{\draw[red,solid,line width=1.0pt](0,0) -- (5mm,0);}}}

\DeclareRobustCommand{\bluedashdotted}{\raisebox{2pt}{\tikz{\draw[blue,dashdotted,line width=1.0pt](0,0) -- (5mm,0);}}}

\DeclareRobustCommand{\greendashed}{\raisebox{2pt}{\tikz{\draw[green,longdashed,line width=1.0pt](0,0) -- (5mm,0);}}}
\DeclareRobustCommand{\magentadashed}{\raisebox{2pt}{\tikz{\draw[magenta2,dashed,line width=1.0pt](0,0) -- (5mm,0);}}}
\DeclareRobustCommand{\blackdashdotdot}{\raisebox{2pt}{\tikz{\draw[black,dashdotdot,line width=1.0pt](0,0) -- (5mm,0);}}}

\title{Multi-agent reinforcement learning for wall modeling in LES of flow over periodic hills}

\shorttitle{MARL for wall modeling in LES of flow over periodic hills}

\author{D. Zhou\footnote[1]{Graduate Aerospace Laboratories, California Institute of Technology},  M. P. Whitmore, K. P. Griffin \and H. J. Bae$\dagger$}

\shortauthor{Zhou et~al.}

\begin{document}
\pagenumbering{gobble}

\setcounter{page}{1}

\maketitle

We develop a wall model for large-eddy simulation (LES) that takes into account various pressure-gradient effects using multi-agent reinforcement learning (MARL). The model is trained using low-Reynolds-number flow over periodic hills with agents distributed on the wall along the computational grid points. The model utilizes a wall eddy-viscosity formulation as the boundary condition, which is shown to provide better predictions of the mean velocity field, rather than the typical wall-shear stress formulation. Each agent receives states based on local instantaneous flow quantities at an off-wall location, computes a reward based on the estimated wall-shear stress, and provides an action to update the wall eddy viscosity at each time step. The trained wall model is validated in wall-modeled LES (WMLES) of flow over periodic hills at higher Reynolds numbers, and the results show the effectiveness of the model on flow with pressure gradients. The analysis of the trained model indicates that the model is capable of distinguishing between the various pressure gradient regimes present in the flow.\\

\hrule

\section{Introduction}

LES is an essential technology for the simulation of turbulent flows. The basic premise of LES is that energy-containing and dynamically important eddies must be resolved everywhere in the domain. This requirement is hard to meet in the near-wall region, as the stress-producing eddies become progressively smaller. The cost involved in resolving the near-wall region has motivated the development of WMLES, which utilizes a wall model to account for the effect of the energetic near-wall eddies. Because of such characteristics, wall modeling has been accepted as the next step to enable the increased use of high-fidelity LES in realistic engineering and geophysical applications.

Recently, \cite{bae2022scientific} demonstrated the efficacy of MARL as a development tool for wall models in zero-pressure-gradient flat-plate flow. In that work, a series of learning agents are distributed along the computational grid points, with each agent receiving local states and rewards then providing local actions at each time step. The MARL model is able to achieve these results by training in situ on moderate-Reynolds-number flows with a reward function only based on the mean wall-shear stress. Therefore, the trained models do not require any higher-fidelity simulation data and are data efficient, unlike supervised learning. Furthermore, MARL can develop novel models that are optimized to accurately reproduce the quantities of interest by discovering dominant patterns in flow physics rather than using \emph{a-priori} parameterization.

In the present study, we extend the methodology of \cite{bae2022scientific} for pressure-gradient flows and apply it to the flow over periodic hills, training on low-Reynolds-number cases and testing on higher-Reynolds-number flows. Our objective is to develop a wall model for LES based on MARL that is robust to pressure-gradient effects in a data-efficient way. The remainder of this report is organized as follows. In Section~\ref{sec:method}, the setup of the flow simulation and reinforcement learning are introduced. In addition, an investigation of two candidate wall boundary conditions is presented. In Section~\ref{sec:results}, tests of the wall model are presented for flow over periodic hills at various Reynolds numbers. Finally, conclusions are drawn in Section~\ref{sec:con}. 

\section{Methodology}
\label{sec:method}

\subsection{Flow simulation}


For the flow solver, we utilize a finite volume, unstructured-mesh incompressible LES code \citep{you2008discrete}. The subgrid-scale (SGS) stress is modeled using the dynamic Smagorinsky model \citep{germano1991dynamic,lilly1992}. 

For training the wall model, a flow configuration that (i) has widely available wall-shear stress profiles for several Reynolds numbers and (ii) does not require tuning of the inlet profile or other boundary conditions is chosen. The flow over periodically arranged hills in a channel \citep{mellen2000large} has well-defined boundary conditions, can be computed at affordable costs, and constitutes a canonical smooth-body separation and reattachment, which is representative of many aerodynamic applications. Furthermore, the flow does not require reconfiguration of the inlet boundary condition for different grid resolutions and wall models, which may be necessary for nonperiodic flows. This configuration has become a popular benchmark test case for validating computational fluid dynamics codes, since numerous experimental and high-fidelity numerical references exist \citep[e.g.,][]{krank2018direct, gloerfelt2019large} that provide extensive data on a wide range of Reynolds numbers covering $700\le Re_H \le 37000$, where $Re_H = U_b H / \nu$, $H$ is the hill height, and $U_b$ is the bulk velocity at the top of the hill. The periodic-hill configuration has the dimensions of $9H\times3.035H\times4.5H$ in the streamwise ($x$), vertical ($y$), and spanwise ($z$) directions, respectively. In the simulations of the present study, periodic boundary conditions are applied in streamwise and spanwise boundaries, and the equilibrium stress-balance wall model \citep[EQWM;][]{kawai2012wall} is employed at the top wall. To maintain constant bulk velocity in time, the flow is driven by time-varying body force  \citep{balakumar2014dns}. A set of three meshes with increasing resolution are used in the present study. More details of the grids utilized are listed in Table~\ref{table1}. The grid is evenly spaced in the $z$ direction and is slightly non-uniform in both the $x$ and $y$ directions. A CFL number of 1 is used for all simulations.

\begin{table}
\begin{center}
\vskip 0.0in
\begin{tabular}{l c r r}
\toprule
Case & Mesh size ($N_x\times N_y\times N_z$) & $x_{sep}/H$ & ${x_{rea}/H}$ \\
\midrule
RLWM & $128\times64\times64$ & -0.03 & 3.91 \\
RLWM, coarse mesh & $64\times32\times32$ & 0.36 & 4.16 \\ 
RLWM, fine mesh & $128\times128\times72$ & 0.21 & 3.85 \\
EQWM & $128\times64\times64$ & 0.57 & 3.05 \\ 
DNS \citep{krank2018direct} & $896\times448\times448$ & 0.20 & 4.51 \\ 
WRLES \citep{gloerfelt2019large} & $512\times256\times256$ & -0.11 & 4.31 \\
\bottomrule
\end{tabular}
\caption{Simulation cases in comparison to reference data, including mesh size, mean separation location $x_{sep}$, and
mean reattachment location $x_{rea}$ at $Re_H=10595$. \label{table1}}
\end{center}
\end{table}

\subsection{Reinforcement learning architecture}

The MARL architecture of wall-model training in the study is based on \cite{bae2022scientific}. During model training, the agents distributed above the wall receive states based on local instantaneous flow information and a reward based on the estimated wall-shear stress, then provide local actions to update the wall boundary condition at each time step. The agents infer a single optimized policy through their repeated interactions with the flow field to maximize its cumulative long-term reward. To utilize MARL as a model development tool for wall models, an open-source reinforcement learning (RL) toolbox smarties \citep{novati2019a} is coupled with the flow solver. The coupling is validated using the same training configuration as the preliminary study of \cite{bae2022scientific}.

\subsection{Boundary conditions for wall modeling}

Previous studies show that how the wall-shear stress is applied affects the mean quantities \citep{balakumar2014dns}. We test two formulations of the boundary condition, namely the wall-shear stress and the wall eddy-viscosity formulations, to determine the appropriate action for the wall model. Two simulations for the flow over periodic hills at $Re_H=10595$ are conducted using the same mesh ($N_x\times N_y\times N_z=128\times64\times64$) but with the two different boundary conditions. For the wall-shear stress formulation, the wall-shear stress $\tau_w$ is set to the mean wall-shear stress from WRLES~\citep{gloerfelt2019large}. For the wall eddy-viscosity formulation, the wall-shear stress is calculated as $\tau_w=(\nu+\nu_{t,w}) (\partial{u_s}/\partial{n})_w$, where ${u_s}$ is the wall-parallel velocity, $n$ denotes the wall-normal direction pointing toward the interior, and the subscript $w$ denotes quantities evaluated at the wall. The wall eddy viscosity $\nu_{t,w}(x)$ is obtained from the interpolated mean-flow field and mean skin friction of WRLES~\citep{gloerfelt2019large}. Similar boundary conditions for wall modeling have been used in the past~\citep{bae2021effect}.

\begin{figure}
\centering
\includegraphics[width=\textwidth]{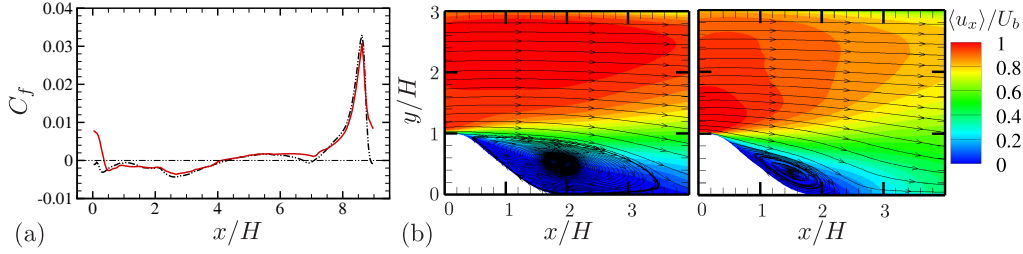}
\caption{(a) Mean skin friction coefficient along the bottom wall at $Re_H=10595$; {\redsolid}, LES using wall eddy-viscosity formulation; \blackdashdotdot, LES using wall-shear stress formulation. (b) Contours of the mean streamwise velocity and the streamlines at $Re_H=10595$: (left) LES using wall eddy-viscosity formulation and (right) LES using wall-shear stress formulation.}
\label{fig:BC_test}
\end{figure}

The mean skin friction coefficients of the two simulations are compared in Figure~\ref{fig:BC_test}(a). The skin friction coefficient is defined as $C_f=\langle \tau_w \rangle/(0.5\rho U_b^2)$, where $\rho$ is the density and $\langle\cdot\rangle$ denotes the averaging in both time and spanwise direction. The $C_f$ of the wall-shear stress formulation is equal to that from WRLES \citep{gloerfelt2019large} by design, whereas the wall eddy-viscosity formulation deviates slightly, with some differences observed near the hill peak and on the leeward side of the hill. Despite the similar predictions of $C_f$, the velocity fields of the two simulations are different [Figure~\ref{fig:BC_test}(b)]. The size of the separation bubble behind the hill is largely underpredicted in the wall-shear stress formulation, whereas the prediction from the wall eddy-viscosity formulation agrees reasonably well with the WRLES results (see Figure~\ref{fig:streamline_Re10595}). 

The underprediction of the separation bubble in the wall-shear stress formulation can be explained by the directional inconsistency of the $u_s$ at the wall-adjacent cell and $\tau_w$. When the direction of the $\tau_w$ is different from $u_s$ at the wall-adjacent cell, the forces acting on the control volume of the wall-adjacent cell will push the $u_s$ in the incorrect direction in a positive feedback loop, leading to an inaccurate prediction of the separation bubble. We find that the wall eddy-viscosity formulation is a more robust method for WMLES of such flows and will utilize it for the remainder of the paper.

\subsection{Training of the wall model}

The model training is carried out based on the LES of periodic-hill channel flow at $Re_H=10595$ using the baseline mesh ($N_x\times N_y\times N_z=128\times64\times64$). A total of 512 agents are uniformly distributed along the bottom wall, and the wall-normal locations of the agents $h_m$ are randomly selected between $0.01H$ and $0.09H$ at each agent location. For each agent, we set three local flow quantities as states: (i) the instantaneous wall-parallel velocity $u_s$; (ii) the wall-normal location of the agent $y_n=h_m$; and (iii) the turbulence strain rate $S_{12}= (\partial{u_s}/\partial{n}+\partial{u_n}/\partial{s})/2$, where $s$ denotes the wall-parallel direction pointing toward the positive $x$-direction. Moreover, to increase the applicability of the wall model for a wide range of flow parameters, the states are nondimensionalized using kinematic viscosity $\nu$ and the instantaneous composite friction velocity $u_{\tau p}=({u_{\tau}^2+u_{p}^2})^{1/2}$, where $u_p=|(\nu/\rho)( \partial{p_w}/\partial{s})|^{1/3}$, $p_w$ is the pressure on the bottom wall, and $u_{\tau}$ is the friction velocity based on the instantaneous modeled wall-shear stress. Nondimensionalized quantities are denoted by superscript $*$. Each agent acts to adjust the local wall eddy viscosity $\nu_{t,w}$ at each time step through a multiplication factor $\nu_{t,w}(t_{i+1}) = a\nu_{t,w}(t_i)$, where $a\in \left[1-\alpha\Delta T U_b/\Delta x,1+\alpha\Delta T U_b/\Delta x\right]$, $\Delta T$ is the action time step, $\Delta x$ is the grid size in $x$ direction, and $\alpha$ is a constant that is selected to be 0.001. The local wall-shear stress can be calculated by the formula $\tau_{w}=\nu (1+\nu_{t,w}^{*})(\partial{u_s}/\partial{n})_w$. The reward $r$ is calculated based on $r(t_{i})=[|\tau_w^{ref}-\tau_w(t_i)|-|\tau_w^{ref}-\tau_w(t_{i-1})|]/\tau_{w,rms}^{ref}$ at each location, where $\tau_w^{ref}$ and $\tau_{w,rms}^{ref}$ are the mean and root-mean-square wall-shear stress from reference simulation, respectively. The reward is proportional to the improvement in the modeled wall-shear stress compared to the one obtained in the previous time step, and an extra reward of 0.1 is added when the modeled $\tau_w$ is within $10\%$ of the reference value. 

During model training, each iteration is initialized with the normalized wall eddy viscosity $\nu_{t,w}^*$ that is randomly selected from $(0,6]$. To generate the initial condition for training, the simulation is started from a flow field generated by the EQWM and run with the given initial $\nu_{t,w}$ for 20 flow-through times (FTTs) to remove numerical artifacts. Each iteration of the model training simulation is conducted for 5 FTTs, and to increase the data efficiency, the $\nu_{t,w}$ is updated every 100 time steps.

The model policy is learned based on a neural network with two hidden layers of 128 units each with a soft sign activation function. The parameters of the neural network are identical to those used in \cite{bae2022scientific}. The model training is advanced for 1.6 million policy gradient steps.

\section{Results}
\label{sec:results}

\subsection{Testing for flow over periodic hills at $\textnormal{Re}_\textnormal{H}=10595$}

To evaluate the performance of the trained wall model (RLWM), three simulations for the periodic-hill channel flow at $Re_H=10595$ are carried out using meshes with different resolutions. The details of the simulation cases are listed in Table~\ref{table1}. The number of agents on the bottom wall is consistent with the number of mesh cells on the wall. The wall-normal matching location of the agents was chosen to be within the first off-wall cell. The wall eddy viscosity $\nu_{t,w}$ is updated based on the model action at every time step. All simulations are run for about 60 FTTs after transients. The flow statistics of all simulations are averaged over spanwise direction and time. For comparison, an LES using EQWM is conducted and results from two high-fidelity reference databases for this flow \citep{krank2018direct,gloerfelt2019large} are included.
 
\begin{figure}
\centering
\includegraphics[width=\textwidth]{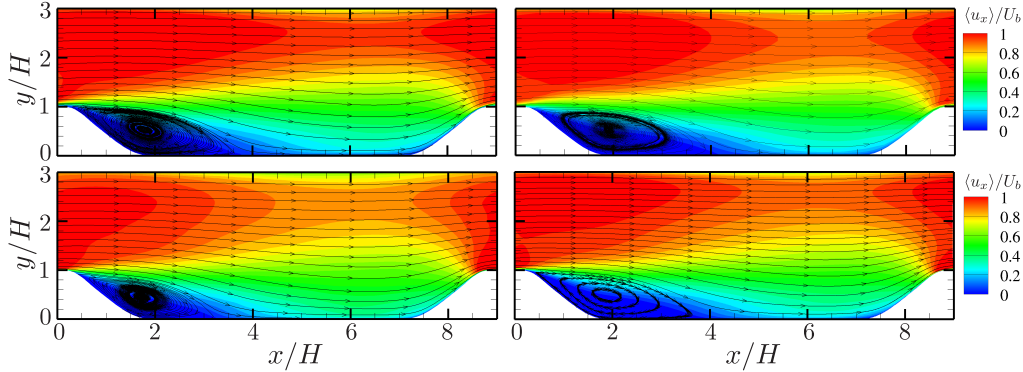}
\caption{Contours of the velocity in $x$ direction and the streamlines at $Re_H=10595$: (top left) RLWM; (top right) RLWM, coarse mesh; (bottom left) EQWM; and (bottom right) WRLES \citep{gloerfelt2019large}.}
\label{fig:streamline_Re10595}
\end{figure}  

Figure~\ref{fig:streamline_Re10595} shows the contours of the mean velocity in $x$ direction and the mean-flow streamlines. The flow separates on the leeward side of the hill due to a strong adverse pressure gradient (APG), and a shear layer is generated near the top of the hill. The flow reattaches in the middle section of the channel, and as the flow approaches the windward side of the downstream hill, it is subjected to a strong favorable pressure gradient (FPG) and accelerates rapidly. The simulations with RLWM successfully capture the separation bubble on the leeward side of the hill and yield more accurate results than the EQWM (see Table~\ref{table1} for quantitative comparison).

\begin{figure}
\includegraphics[width=\textwidth]{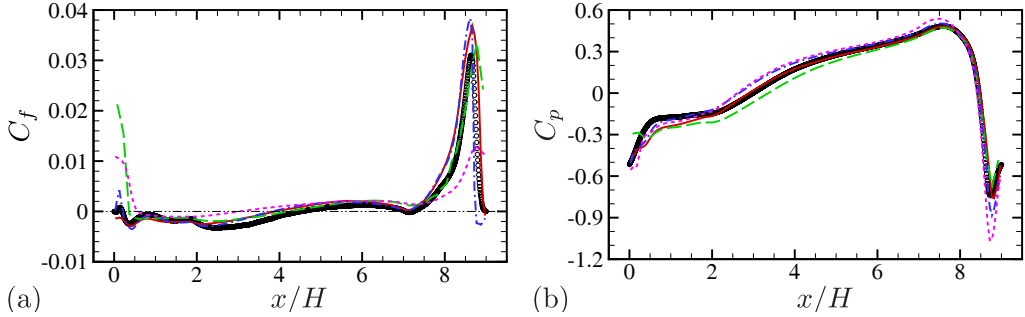}
\caption{(a) Mean skin friction coefficient and (b) mean pressure coefficient along the bottom wall at $Re_H=10595$. Lines indicate {\redsolid}, RLWM; {\greendashed}, RLWM, coarse mesh; {\bluedashdotted}, RLWM, fine mesh; {\magentadashed}, EQWM; $\circ$, DNS~\citep{krank2018direct}.}
\label{fig:CfCp_Re10595}
\end{figure}

The predictions of $C_f$ and the mean pressure coefficient $C_p$ are shown in Figure~\ref{fig:CfCp_Re10595}. The mean pressure coefficient is defined as $C_p=(\langle p_w\rangle-\langle p_{ref}\rangle)/(0.5\rho U_b^2)$, where the pressure at $x/H=0$ on the top wall is chosen as reference pressure $p_{ref}$ \citep{krank2018direct}. Regarding $C_f$, the results from RLWM simulations are in reasonable agreement with the direct numerical simulation (DNS) data, with large deviations found only near the top of the hill where the skin friction reaches its maximum value and then rapidly decreases to a negative value. As the mesh resolution increases, it yields a more accurate prediction of skin friction on the leeward side, but the maximum value of the skin friction on the windward side of the hill still is overpredicted. However, the results are better than the EQWM simulation, which largely underpredicts the skin friction on the windward side and overpredicts the skin friction on the leeward side. Furthermore, the mean locations of the separation and reattachment points (listed in Table~\ref{table1}) are better predicted by the RLWM, consistent with the streamline shown in Figure~\ref{fig:streamline_Re10595}. Additionally, note that the separation bubble size changes as the mesh resolution increases or decreases from the baseline, and more details about the velocity field will be discussed later in this section. 

\begin{figure}
\centering
\includegraphics[width=0.62\textwidth]{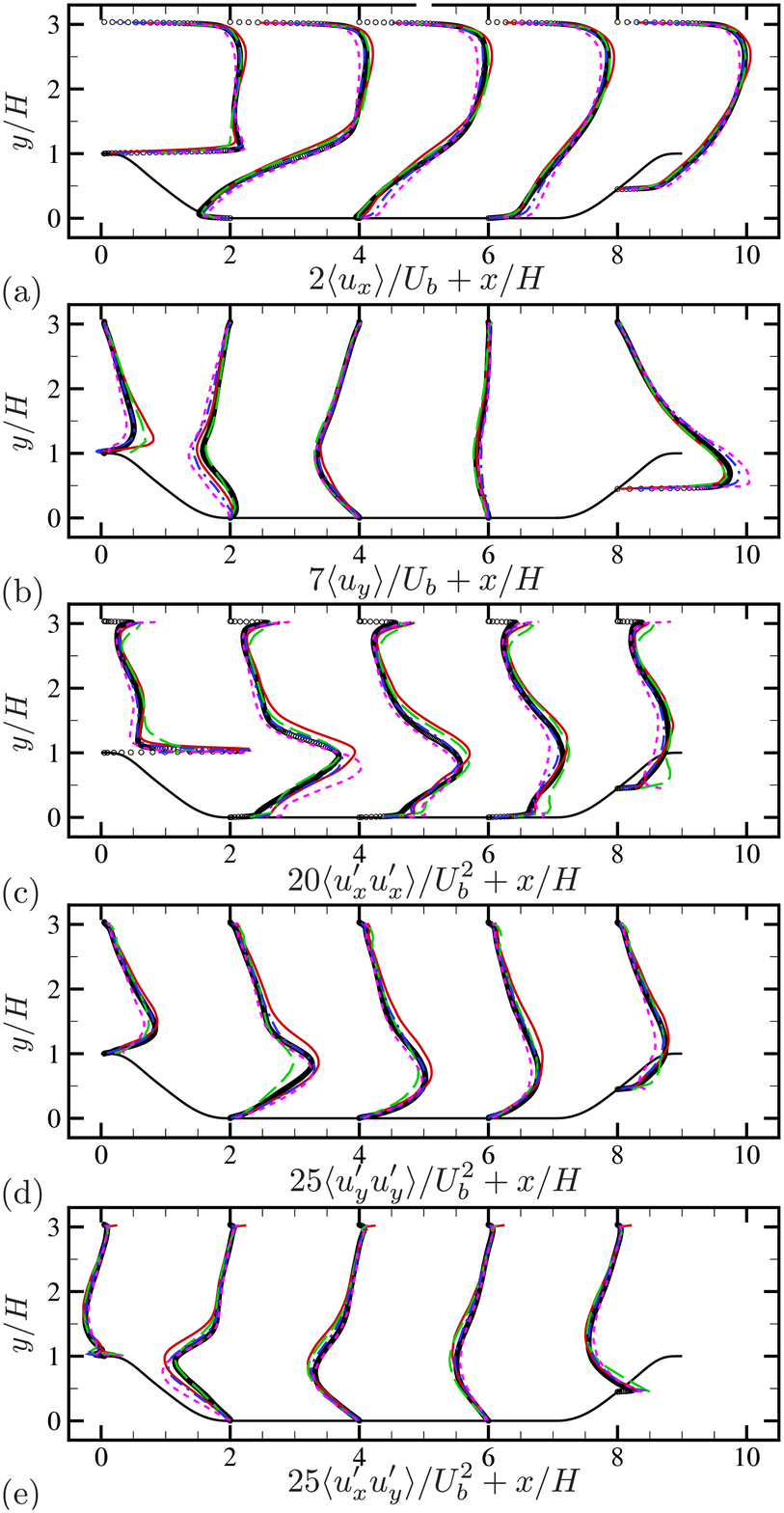}
\vspace{0.3cm}
\caption{ Mean velocity and Reynolds stress components at $Re_H=10595$: (a) streamwise velocity, (b) vertical velocity, (c) streamwise Reynolds stress, (d) vertical Reynolds stress, and (e) Reynolds shear stress. Lines indicate {\redsolid}, RLWM; {\greendashed}, RLWM, coarse mesh; {\bluedashdotted}, RLWM, fine mesh; {\magentadashed}, EQWM; $\circ$, DNS~\citep{krank2018direct}.}
\label{fig:flowstates_Re10595}
\end{figure} 

All simulations capture the qualitative trend of the mean $C_p$ on the bottom wall including the APG and FPG regimes, but relatively larger deviations among the simulation cases are visible near the top of the hill ($x/H\geq8.5$ or $x/H\leq0.5$) where the pressure sees a sudden change from strong FPG to strong APG and the flow separation emerges. Overall, the RLWM provides more accurate predictions of $C_f$ and $C_p$ than the EQWM.

Quantitative comparison of mean velocity and Reynolds stress components at five streamwise locations ($x/H=0.05, 2, 4, 6, \text{ and } 8$) are shown in Figure~\ref{fig:flowstates_Re10595}. The results of the mean velocity profiles and Reynolds stress profiles from the RLWM simulations agree reasonably well with each other and with the reference DNS data. The discrepancies are visible for the regions near channel center. Note that the prediction of the velocity field depends not only on the wall boundary conditions but also on the SGS model, and the varying performance of the SGS model in different meshes may contribute to the inconsistency of those three RLWM simulations. Furthermore, the prediction from EQWM simulation is less ideal, particularly for the mean streamwise velocity $\langle u_x\rangle/U_b$.

\begin{figure}
\centering
\includegraphics[width=\textwidth]{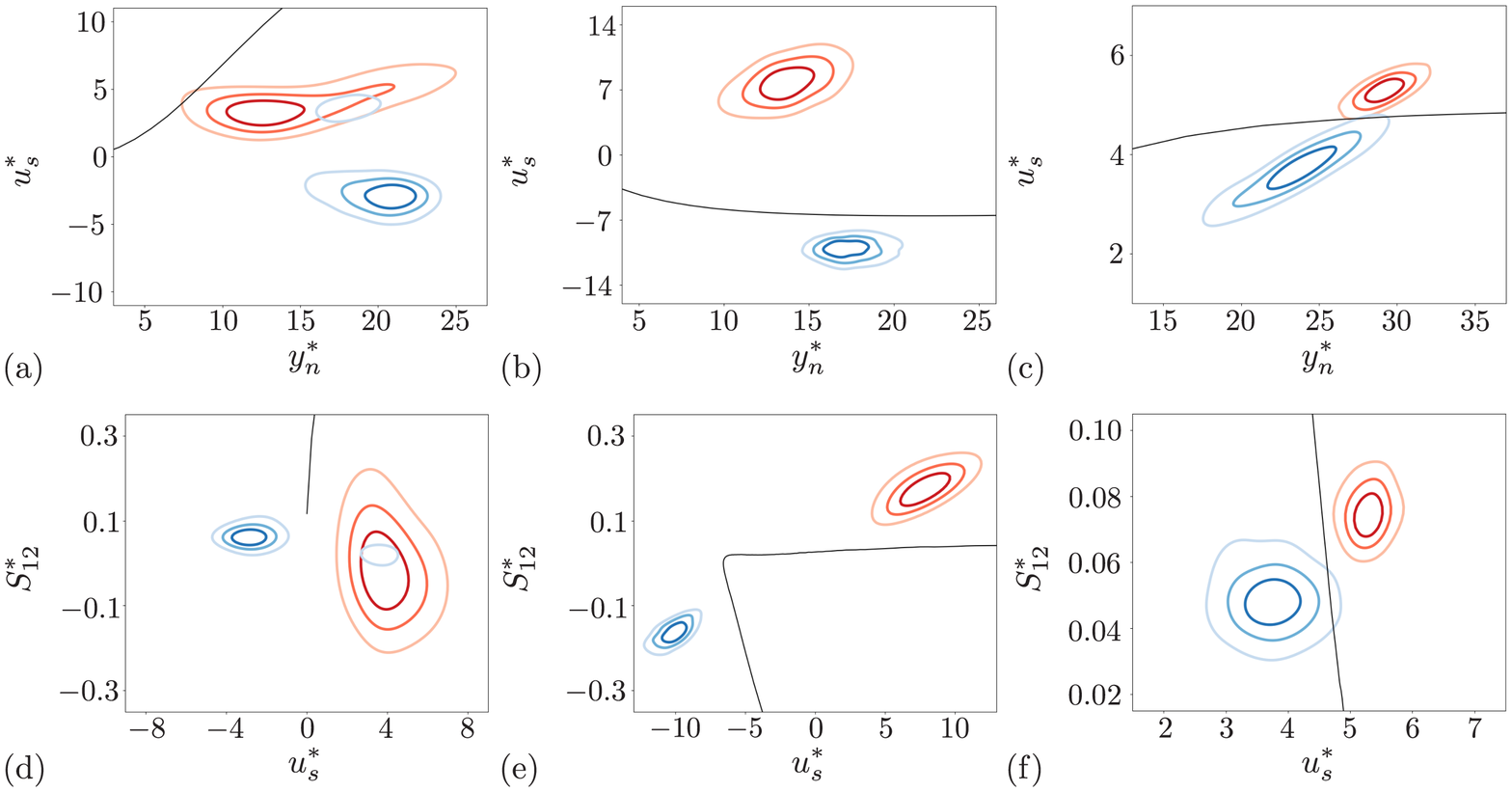}
\caption{PDFs of states for RLWM conditioned to events with $r>0.1$ and $a<0.9996$ (blue) or $a>1.0004$ (red) at (a,d) $x/H=0.025$, (b,e) $x/H=2$, and (c,f) $x/H=8.3$ at $Re_H = 10595$: PDFs of states (a--c) $u_s^*$ and $y_n^*$ and (d--f) $S_{12}^*$ and $u_s^*$. The black line is the mean relation of the states from WRLES~\citep{gloerfelt2019large}. Contour levels are $25$, $50$, and $75\%$ of the maximum value.}
\label{fig:state_act}
\end{figure}

To better understand the mechanism of the trained model, we examine the state-action maps, which are the probability density functions (PDFs) of the likelihood that the model takes a particular action conditioned on the occurrence of positive rewards. Figure~\ref{fig:state_act} shows the maps based on instantaneous states and actions at three streamwise positions $x/H=0.025, 2, \text{ and } 8.3$ which are located at the top of the hill, within the separation bubble, and the windward side of the hill, respectively. Overall, the action contour lines for increasing and decreasing $\nu_{t,w}$ are well separated with respect to the mean quantities obtained from the WRLES simulation~\citep{gloerfelt2019large}, which illustrates that the model is able to distinguish flow states and provide appropriate actions. Some minor overlaps of the contour lines not clearly separated by the mean relation from WRLES are visible at the top of the hill [see Figures~\ref{fig:state_act}(a,d)] where the APG is strong. This implies that the capability of the trained model to robustly adjust the wall-shear stress through appropriate actions is less ideal within this regime.

\subsection{Testing for flow over periodic hills at higher Reynolds numbers}

In this section, the RLWM is applied to a WMLES of periodic-hill channel flow at $Re_H=19000$ and $37000$. The simulations are conducted by using the baseline mesh ($128\times64\times64$) and the coarse mesh ($64\times32\times32$), and the implementation of RLWM is similar to the simulations at $Re_H=10595$. All simulations are run for about 60 FTTs after transients. The results from EQWM with baseline mesh and the WRLES \citep{gloerfelt2019large} are included for comparison. 

\begin{figure}
\centering
\includegraphics[width=\textwidth]{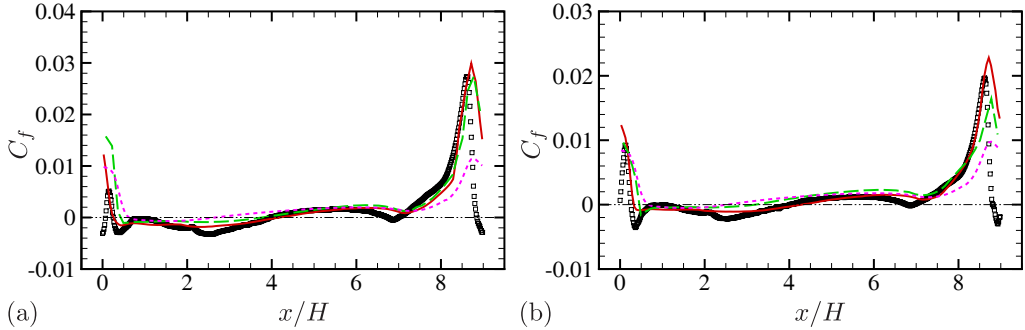}
\caption{Mean skin friction coefficient along the bottom wall at (a) $Re_H=19000$ and (b) $Re_H=37000$. Lines indicate {\redsolid}, RLWM; {\greendashed}, RLWM, coarse mesh; {\magentadashed}, EQWM; \square, WRLES~\citep{gloerfelt2019large}.}
\label{fig:Cf_higherRe}
\end{figure}

The mean skin friction coefficients along the bottom wall at $Re_H=19000$ and $37000$ are shown in Figure~\ref{fig:Cf_higherRe}(a,b). The distributions of $C_f$ at higher Reynolds numbers have a similar shape as the one shown in Figure~\ref{fig:CfCp_Re10595}(a). As the Reynolds number increases, the maximum value of the $C_f$ on the windward side of the hill decreases. The results from the two RLWM simulations are in reasonable agreement with the WRLES results. Regarding the separation point, the predicted locations from the RLWM simulations are further downstream than that of the WRLES. The predicted reattachment location from the RLWM simulation with baseline mesh is slightly downstream from that of the WRLES, but the prediction from the RLWM simulation with coarse mesh is further upstream. Compared to the EQWM simulation, where the separation bubble size is largely underpredicted and the maximum value of $C_f$ on the windward side of the hill is much smaller than the value of WRLES, RLWM is more accurate.  

\begin{figure}
\centering
\includegraphics[width=0.62\textwidth]{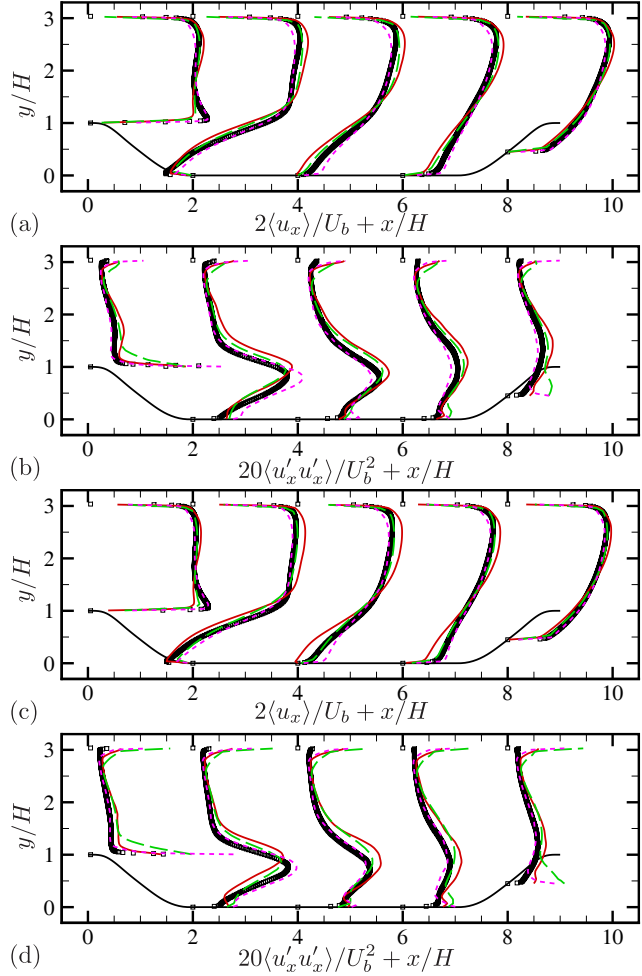}
\caption{Mean velocity and Reynolds stress components at (a,b) $Re_H = 19000$ and (c,d) $Re_H=37000$: (a,c) mean streamwise velocity and (b,d) streamwise Reynolds stress. Lines indicate {\redsolid}, RLWM; {\greendashed}, RLWM, coarse mesh; {\magentadashed}, EQWM; \square, WRLES~\citep{gloerfelt2019large}.}
\label{fig:flowstates_higherRe}
\end{figure}

Figure~\ref{fig:flowstates_higherRe} shows the profiles of streamwise components of mean velocity and Reynolds stress at five streamwise stations ($x/H=0.05, 2, 4, 6, \text{ and } 8$) for $Re_H=19000$ and $37000$. The deviations among the profiles from RLWM simulations and the WRLES profiles grow as the Reynolds number increases, albeit slightly, which implies the degradation of performance in the RLWM. Particularly for the Reynolds stress, the EQWM outperforms the RLWM even though the prediction of $C_f$ is inaccurate, as shown in Figure~\ref{fig:Cf_higherRe}. Furthermore, at $Re_H=19000$ the coarse-mesh RLWM simulation results agree better with the WRLES results than the baseline results. However, the performance of the SGS model affects the velocity field away from the wall more than the wall model, and this could be an indication that a better SGS model is necessary.  

\section{Conclusions}
\label{sec:con}

In this work, a wall model that can adapt to various pressure-gradient effects is developed for turbulent flow over periodic hills using multi-agent reinforcement learning. The model behaves as a control policy for wall eddy viscosity to predict the correct wall-shear stress. The optimized policy of the wall model is learned through the training process based on LES of low-Reynolds-number flow over periodic hills with cooperating agents using the recovery of the correct wall-shear stress as a reward. 

The developed wall model is first validated in the LES of the periodic hill configuration at the same Reynolds number of model training. The wall model provides good predictions of mean wall-shear stress, mean wall pressure, and the mean velocity as well as Reynolds stress in the flow field. The test results also show that the developed model outperforms the EQWM. The performance of the developed model is further evaluated at $Re_H=19000$ and $37000$. Good predictions are obtained for the mean wall-shear stress. 

Overall, we have extended the MARL-based wall modeling paradigm and demonstrated that it is robust in flows with pressure gradients. Additional investigations into the generalizability of the developed wall model for different geometries and flow configurations need to be carried out, and the grid sensitivity of the model needs to be further analyzed.

\subsection*{Acknowledgments}
The authors acknowledge the XSEDE computing resource from PSC under project PHY210119. The authors would like to thank Dr.~Xiang Yang for their insightful remarks on the report.

\bibliographystyle{ctr}

\begin{thebibliography}{12}
\expandafter\ifx\csname natexlab\endcsname\relax\def\natexlab#1{#1}\fi

\bibitem[Bae \& Koumoutsakos(2022)]{bae2022scientific}
{\sc Bae, H.~J. \& Koumoutsakos, P.} 2022 Scientific multi-agent reinforcement
  learning for wall-models of turbulent flows. {\em Nat. Commun.\/} {\bf
  13}, 1443.

\bibitem[Bae \& Lozano-Dur{\'a}n(2021)]{bae2021effect}
{\sc Bae, H.~J. \& Lozano-Dur{\'a}n, A.} 2021 Effect of wall boundary
  conditions on a wall-modeled large-eddy simulation in a finite-difference
  framework. {\em Fluids\/} {\bf 6}, 112.

\bibitem[Balakumar {\em et~al.\/}(2014)Balakumar, Park \&
  Pierce]{balakumar2014dns}
{\sc Balakumar, P., Park, G. \& Pierce, B.} 2014 \uppercase{DNS},
  \uppercase{LES}, and wall-modeled \uppercase{LES} of separating flow over
  periodic hills. {\em Proceedings of the Summer Program\/}, Center for Turbulence Research, Stanford University, pp.~407--415.

\bibitem[Germano {\em et~al.\/}(1991)Germano, Piomelli, Moin \&
  Cabot]{germano1991dynamic}
{\sc Germano, M., Piomelli, U., Moin, P. \& Cabot, W.~H.} 1991 A dynamic
  subgrid-scale eddy viscosity model. {\em Phys. Fluids\/} {\bf 3},
  1760--1765.

\bibitem[Gloerfelt \& Cinnella(2019)]{gloerfelt2019large}
{\sc Gloerfelt, X. \& Cinnella, P.} 2019 Large eddy simulation requirements for
  the flow over periodic hills. {\em Flow Turbul. Combust.\/} {\bf 103},
  55--91.

\bibitem[Kawai \& Larsson(2012)]{kawai2012wall}
{\sc Kawai, S. \& Larsson, J.} 2012 Wall-modeling in large eddy simulation:
  Length scales, grid resolution, and accuracy. {\em Phys. Fluids\/} {\bf
  24}, 015105.

\bibitem[Krank {\em et~al.\/}(2018)Krank, Kronbichler \& Wall]{krank2018direct}
{\sc Krank, B., Kronbichler, M. \& Wall, W.~A.} 2018 Direct numerical
  simulation of flow over periodic hills up to $\text {Re}_{H}=10{,}595$. {\em
  Flow Turbul. Combust.\/} {\bf 101}, 521--551.

\bibitem[Lilly(1992)]{lilly1992}
{\sc Lilly, D.~K.} 1992 A proposed modification of the Germano subgrid-scale
  closure method. {\em Phys. Fluids A\/} {\bf 4}, 633--635.

\bibitem[Mellen {\em et~al.\/}(2000)Mellen, Fr{\"o}hlich \&
  Rodi]{mellen2000large}
{\sc Mellen, C.~P., Fr{\"o}hlich, J. \& Rodi, W.} 2000 Large eddy simulation of
  the flow over periodic hills. In {\em Proceedings of the 16th IMACS World Congress\/}, pp.
  21--25.

\bibitem[Novati \& Koumoutsakos(2019)]{novati2019a}
{\sc Novati, G. \& Koumoutsakos, P.} 2019 Remember and forget for experience
  replay. In {\em Proceedings of the 36th International Conference on Machine
  Learning\/}, pp.
  4851--4860.


\bibitem[You {\em et~al.\/}(2008)You, Ham \& Moin]{you2008discrete}
{\sc You, D., Ham, F. \& Moin, P.} 2008 Discrete conservation principles in
  large-eddy simulation with application to separation control over an airfoil.
  {\em Phys. Fluids\/} {\bf 20}, 101515.
\end{thebibliography}

\end{document}